\documentclass[aps,superscriptaddress,prc,twocolumn,nofootinbib]{revtex4}

\usepackage{graphicx}
\usepackage{epstopdf}
\usepackage{subfigure}
\usepackage{epsfig}
\usepackage{amsmath,amssymb,amsfonts}
\usepackage{color}
\usepackage[utf8]{inputenc}
\allowdisplaybreaks
\usepackage[bookmarks,
                bookmarksopen = true,
                bookmarksnumbered = true,
                linktocpage,
                colorlinks = true,
                linkcolor = blue,
                urlcolor  = blue,
                citecolor = blue,
                anchorcolor = green,
                hyperindex = true,
                hyperfigures]
                {hyperref}
\usepackage{multirow}
\usepackage{array}

\begin{document}

\title{Flavor Hierarchy of Jet Energy Correlators inside the Quark-Gluon Plasma}

\author{Wen-Jing Xing}
\affiliation{Institute of Frontier and Interdisciplinary Science, Shandong University, Qingdao, Shandong 266237, China}

\author{Shanshan Cao}
\email{shanshan.cao@sdu.edu.cn}
\affiliation{Institute of Frontier and Interdisciplinary Science, Shandong University, Qingdao, Shandong 266237, China}

\author{Guang-You Qin}
\email{guangyou.qin@ccnu.edu.cn}
\affiliation{Institute of Particle Physics and Key Laboratory of Quark and Lepton Physics (MOE), Central China Normal University, Wuhan, 430079, China}

\author{Xin-Nian Wang}
\email{xnwang@ccnu.edu.cn}
\affiliation{Institute of Particle Physics and Key Laboratory of Quark and Lepton Physics (MOE), Central China Normal University, Wuhan, 430079, China}

\date{\today}


\begin{abstract}

Heavy flavor jets provide ideal tools to probe the mass effect on jet substructure in both vacuum and quark-gluon plasma (QGP).
Energy-energy correlator (EEC) is an excellent jet substructure observable owning to its strong sensitivity to jet physics at different scales.
We perform a complete realistic simulation on medium modification of heavy and light flavor jet EEC in heavy-ion collisions.
A clear flavor hierarchy is observed for jet EEC in both vacuum and QGP due to the mass effect.
The medium modification of inclusive jet EEC at different angular scales exhibits very rich structure: suppression at intermediate angles, and enhancement at small and large angles, which can be well explained by the interplay of mass effect, energy loss, medium-induced radiation and medium response.
These unique features of jet EEC are shown to probe the physics of jet-medium interaction at different scales, and can be readily validated by upcoming experiments.

\end{abstract}

\maketitle


{\it \color{blue} Introduction --} Energetic nuclear collisions conducted at the Relativistic Heavy-Ion Collider (RHIC) and the Large Hadron Collider (LHC) provide a unique opportunity to heat the Quantum Chromodynamics (QCD) vacuum and release the quark degrees of freedom from hadrons, creating a color-deconfined matter called quark-gluon plasma (QGP)~\cite{Gyulassy:2004zy,Jacobs:2004qv,Busza:2018rrf,Elfner:2022iae}.
Jets provide a powerful probe to the dynamics of QGP at different scales~\cite{Wang:1992qdg,Bass:2008rv,Qin:2015srf,Majumder:2010qh,Cao:2020wlm,Cao:2022odi,Cao:2024pxc}.
For example, the global model-to-data comparisons of jet suppression show that the jet transport coefficient inside the QGP is an order of magnitude larger than that inside a cold nucleus~\cite{JET:2013cls,JETSCAPE:2021ehl,Xie:2022ght,Chen:2024epd}.
Intrinsic properties of the QGP, such as its viscosity and equation of state, can also been constrained by the jet observables and data~\cite{Karmakar:2023ity,Liu:2023rfi}.

In recent years, significant attention has been paid to jet substructure~\cite{Chien:2015hda,Casalderrey-Solana:2016jvj,Tachibana:2017syd,KunnawalkamElayavalli:2017hxo,Luo:2018pto,Chang:2017gkt,Mehtar-Tani:2016aco,Milhano:2017nzm,Chen:2020tbl}, $\gamma$/$Z$-triggered hadron distribution~\cite{Qin:2009bk,Chen:2017zte,Zhang:2018urd,Chen:2020tbl,Yang:2021qtl}, and hadron chemistry inside and around the jet cone~\cite{Chen:2021rrp,Luo:2021voy}, which can probe more detailed structures of jets and jet-medium interaction.
A particularly interesting jet substructure observable is the energy correlator due to its sensitivity to the intrinsic and emergent scales in jet physics.
The energy correlator has been extensively discussed in conformal field theories~\cite{Hofman:2008ar}.
Recent studies have shown that an $N$-point energy correlator within jets, defined as $\langle\mathcal{E}(\vec{n}_1)\mathcal{E}(\vec{n}_2)\ldots\mathcal{E}(\vec{n}_N)\rangle$ with $\mathcal{E}(\vec{n}_i)$ denoting the energy flux along the $\vec{n}_i$ direction, can reveal the angular scale associated with the confinement of quarks and gluons into hadrons~\citep{Komiske:2022enw,Tamis:2023guc,CMS:2024mlf}, and also the scale of gluon saturation in high-energy nucleons~\citep{Liu:2022wop,Liu:2023aqb,Cao:2023oef}.
The two-point energy correlator, refereed to as the energy-energy correlator (EEC) below, have recently been extended to heavy-ion collisions.
For example, Refs.~\cite{Andres:2022ovj,Andres:2023xwr} use the BDMPS-Z formalism to study the light quark jets passing through a static medium, and argue that the onset of color coherence in the quark-gluon splitting can manifest in the medium-modified EEC.
Refs.~\cite{Barata:2023zqg,Barata:2023bhh,Andres:2024ksi} further investigate the effects of dynamically evolving medium, the medium anisotropy, parton energy loss and the confinement transition on jet EEC using semi-analytical calculations.
The first realistic analysis of jet EEC based on comprehensive simulations of jet-QGP interactions has been conducted in Ref.~\cite{Yang:2023dwc}, which shows that the medium-induced radiation and medium response can manifest in the EEC of $\gamma$-triggered jets.
In particular, EEC appears to be sensitive to the short-distance structure of the QGP, quantified by Debye screening mass.

Heavy flavor jets, produced from charm and bottom quarks, provide a direct access to study the mass effect on jet substructure in both vacuum and the QGP.
Tremendous effort has been devoted to studying the so-called dead-cone effect, i.e., the suppression of the gluon emission from a massive quark within a forward cone of $\theta_0 \sim m_Q/E$, with $m_Q$ and $E$ being the mass and energy of heavy quarks, respectively.
Recently, the application of jet substructure techniques to heavy flavor jets in $p+p$ collisions has led to the first observation of the dead-cone effect in QCD~\citep{ALICE:2021aqk}.
Note that the mass effect on medium-induced radiation can be indirectly studied from the flavor hierarchy of parton energy loss and jet suppression in heavy-ion collisions~\cite{Uphoff:2011ad,Nahrgang:2013saa,Djordjevic:2013pba,Xing:2019xae,Xing:2023ciw,Zhang:2023oid}.
Recently, the EEC of charm and bottom jets has been studied in $p+p$~\citep{Craft:2022kdo} and A+A~\citep{Andres:2023ymw} collisions, aiming to probe the dead cone effect.
However, the EEC study of heavy flavor jets in $A+A$ collisions has used a simplified ``brick" model and neglected many important contributions to the EEC, such as parton energy loss and medium response in realistic hydrodynamically evolving medium.

In this work, we focus on the flavor hierarchy of jet EEC in heavy-ion collisions, by performing a complete and realistic simulation of jet production and evolution in QGP as described by relativistic hydrodynamics, including the effects from quark mass, parton energy loss, elastic collisions, medium-induced radiation, medium response, as well as the trigger bias (due to jet energy loss) in the jet measurement.
As will be shown later, the interplay of all these effects is crucial to understand the flavor hierarchy of inclusive jet EEC and its unique nuclear modification pattern in $A+A$ collisions, i.e., suppression at intermediate angles, and enhancement at both small and large angles, relative to $p+p$ collisions.
Our study shows that heavy flavor jet EEC is a powerful probe to jet substructure and jet-medium interaction at different scales.

\begin{figure}[tbp!]
\includegraphics[width=0.90\linewidth]{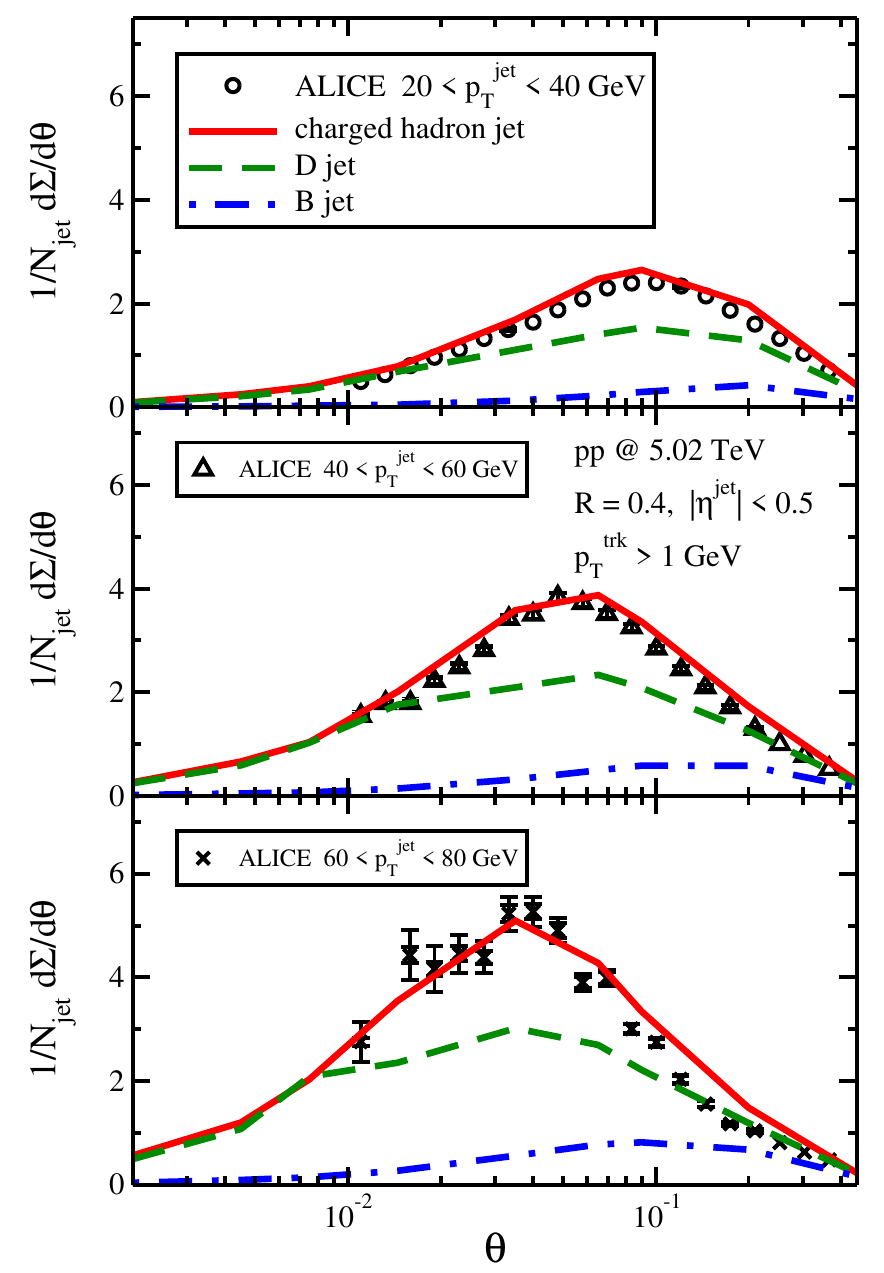}
\caption{(Color online) Angular distributions of the EEC for charged hadron jets, $D$-tagged jets and $B$-tagged jets in $p+p$ collisions at $\sqrt{s_\mathrm{NN}}=5.02$~TeV, for three different $p_{\mathrm{T}}^\mathrm{jet}$ ranges: the upper panel for $(20,40)$~GeV, middle for $(40,60)$~GeV, and lower for $(60,80)$~GeV. The ALICE data on charged hadron jet EEC~\citep{WQF:2023} are shown for comparison.}
\label{fig1}
\end{figure}

{\it \color{blue} Jet EEC's in $p+p$ collisions --} In Fig~\ref{fig1}, we show the EEC spectra $d\Sigma(\theta)/d\theta$ for charged hadron jets, $D$-tagged jets and $B$-tagged jets, averaged over the number of jets ($N_\mathrm{jet}$), with respect to the angle ($\theta$) in $p+p$ collisions at $\sqrt{s_\mathrm{NN}}$=5.02~TeV, obtained from Pythia~8~\cite{Sjostrand:2006za,Sjostrand:2014zea} simulation. Jets within the pseudorapidity region $|\eta^{\mathrm{jet}}|<0.5$ are constructed using tracks with transverse momentum $p_\mathrm{T}>1$~GeV (this kinematics cut also applies to $D$ and $B$ mesons for heavy flavor jets) via the Fastjet package~\citep{Cacciari:2011ma} with the anti-$k_{\rm{T}}$ algorithm~\cite{Cacciari:2008gp} and a cone size of $R=0.4$. The EEC is analyzed using the following formula:
\begin{equation}
\label{eq:defEEC}
\frac{d\Sigma(\theta)}{d\theta} = \frac{1}{\Delta\theta} \sum\nolimits_{|\theta_{ij}-\theta|< \frac{\Delta\theta}{2}} \frac{p_{\mathrm{T},i}(\vec{n}_i)\, p_{\mathrm{T},j}(\vec{n}_j)}{p_{\mathrm{T,jet}}^{2}},
\end{equation}
with $\theta_{ij}$ being the relative angle given by $\cos \theta_{ij} = \vec{n}_{i}\cdot \vec{n}_{j}$ and $\Delta\theta$ the bin size. The sum is performed over all pairs of constituent particles within a jet, with $p_{\mathrm{T,jet}}$ being the total transverse momentum of the jet, $p_{\mathrm{T},i(j)}$ and $\vec{n}_{i(j)}$ the transverse momentum and the direction in the transverse plane of the $i$($j$)-th constituent. The final result is the average over all jets within a desired kinematic region.
From the top to bottom panels, Pythia provides a good description of the ALICE data~\cite{WQF:2023} on the EECs of charged jets within various $p_\mathrm{T}^\mathrm{jet}$ regions.
Comparing different curves in each panel, we observe a clear hierarchy for the EEC distribution tagged by different jet flavors. From charged jets to $D$-jets to $B$-jets, the magnitude of EEC decreases and the peak position of EEC shifts towards larger angle. This result is in line with the mass effect on vacuum parton shower, which suppresses the gluon emission from heavy quarks, especially within the angle $\theta_0 \sim m_Q/E$.

{\it \color{blue} Jet evolution in QGP --} To study nuclear modification of EEC in heavy-ion collisions, jet partons developed from vacuum parton shower in Pythia are fed to the linear Boltzmann transport (LBT) model~\citep{Cao:2016gvr,Luo:2023nsi} for their further evolution inside QGP.
Both elastic and inelastic scatterings between jet partons and medium partons are included in LBT, whose impact on the phase space distribution of jet partons $f_a(\vec{x}_a,\vec{p}_a$,t) is described by the Boltzmann equation as
\begin{equation}
\label{eq:Boltzmann}
	p_a \cdot \partial f_a = E_a\left[C^{\rm el}(f_a) + C^{\rm inel}(f_a\right)].
\end{equation}
Here, $p_a=(E_a,\vec{p}_a)$ denotes the four-momentum of jet parton, $C^{\rm el}$ and $C^{\rm inel}$ are the collision integrals for elastic and inelastic processes, respectively. The elastic scattering rate of a jet parton with medium constituent can be extracted from $C^{\rm el}$ as
\begin{align}
\label{eq:rate}
\Gamma_a^\mathrm{el}&(E_a,T)=\sum_{b,(cd)}\frac{\gamma_b}{2E_a}\int \prod_{i=b,c,d}\frac{d^3p_i}{E_i(2\pi)^3} f_b(E_b,T) \nonumber\\
&\times [1\pm f_c(E_c,T)][1\pm f_d(E_d,T)] S_2(\hat{s},\hat{t},\hat{u})\nonumber\\
&\times (2\pi)^4\delta^{(4)}(p_a+p_b-p_c-p_d)|\mathcal{M}_{ab\rightarrow cd}|^2,
\end{align}
where the sum is over all possible flavors of medium parton $b$ and final state partons $c$ and $d$, $\gamma_b$ is the spin-color degeneracy factor of $b$, $T$ is the medium temperature, $f_i$ ($i=b,c,d$) takes the Bose (Fermi) distribution for gluons (quarks) in the medium rest frame, and $|M_{ab\rightarrow cd}|^2$ is the scattering amplitude square for an $ab\rightarrow cd$ process. To avoid divergence in the matrix element at the leading order, $S_2(\hat{s},\hat{t},\hat{u})=\theta(\hat{s}\ge 2\mu_\mathrm{D}^2)\,\theta(-\hat{s}+\mu^2_\mathrm{D}\le \hat{t} \le -\mu_\mathrm{D}^2)$ is introduced, in which $\hat{s}$, $\hat{t}$, $\hat{u}$ are the Mandelstam variables and $\mu_{\rm{D}}^2=4\pi\alpha_\mathrm{s}T^2(N_c + N_f/2)/3$ is the Debye screening mass, with $N_c$ and $N_f$ the color and flavor numbers, respectively. We assume light flavor quarks and gluons are massless, and use $m_c=1.3$~GeV for charm quarks and $m_b=4.2$~GeV for bottom quarks.

Scatterings that induce gluon bremsstrahlung are categorized as inelastic scatterings. Their rates are related to the number of emitted gluons per unit time as
\begin{equation}
\label{eq:gluonnumber}
\Gamma_a^\mathrm{inel} (E_a,T,t) = \int dzdk_\perp^2 \frac{1}{1+\delta^{ag}}\frac{dN_g^a}{dz dk_\perp^2 dt},
\end{equation}
where the gluon spectrum ${dN_g^a}/(dz dk_\perp^2 dt)$ can be calculated within the higher-twist energy loss formalism~\cite{Wang:2001ifa,Zhang:2003wk,Majumder:2009ge}, $\delta^{ag}$ is to avoid double-counting the $g\rightarrow gg$ rate from its splitting function, $z$ and $k_\perp$ are the fractional energy and transverse momentum of an emitted gluon relative to its parent parton, and $t$ denotes time.

Using the elastic and inelastic scattering rates above, we carry out Monte-Carlo simulation of scatterings between jet partons and medium partons. In the simulation, we track not only jet partons and their emitted gluons, but also thermal partons scattered out of the medium (called ``recoil partons") and the associated energy depletion inside the medium (called ``back-reaction" or ``negative partons"). In our simulation, recoil partons are allowed to continue scattering with medium partons in the same way as jet partons do. The combination of recoil and negative partons constitutes jet-induced medium excitation, or medium response, within the LBT model, which has been shown essential for a quantitative understanding of jet observables, including quenching~\cite{He:2018xjv}, flow~\cite{He:2022evt} and substructure~\cite{Luo:2018pto}. Contribution from negative partons needs to be subtracted from that from regular (positive) ones when evaluating jet observables. For the EEC, $\langle(\mathcal{E}_\mathrm{p}(\vec{n}_1)-\mathcal{E}_\mathrm{n}(\vec{n}_1)(\mathcal{E}_\mathrm{p}(\vec{n}_2)-\mathcal{E}_\mathrm{n}(\vec{n}_2)\rangle$ is calculated, with subscripts p and n denoting positive and negative particles, respectively. The sole parameter in LBT is the strong coupling strength $\alpha_\mathrm{s}$, which affects the magnitudes of $|M_{ab\rightarrow cd}|^2$ and $\mu_\mathrm{D}$ in elastic scatterings, and also enters the bremsstrahlung gluon spectra through parton splitting functions and jet quenching parameter $\hat{q}$~\cite{JET:2013cls}. In this work, we use $\alpha_\mathrm{s}=0.15$, which has provided a good description of jet observables in earlier LBT calculations.

\begin{figure}[tbp]
\includegraphics[width=0.99\linewidth]{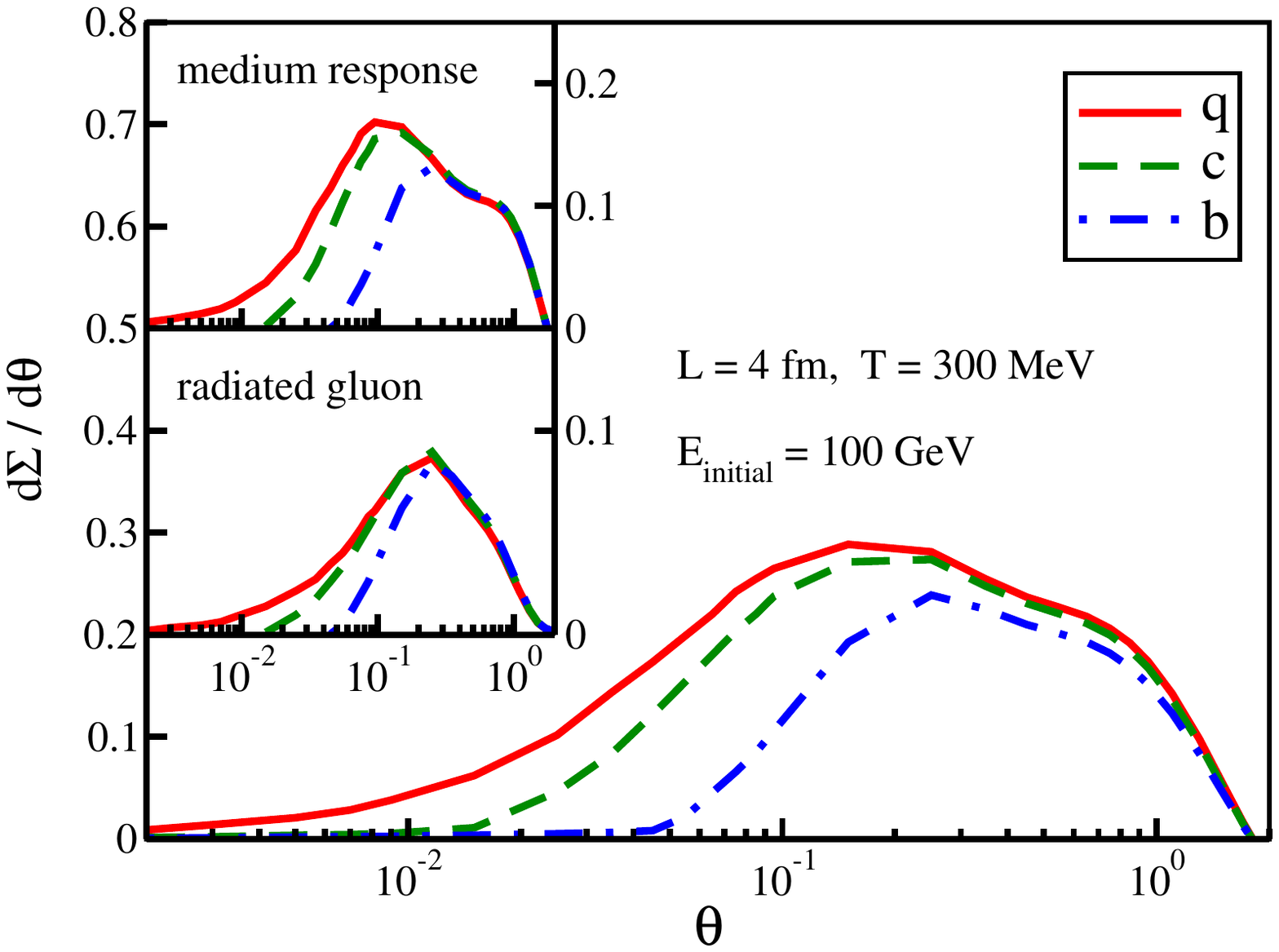}
\caption{(Color online) Angular distributions of EEC for different flavors of jets produced from a single parton with initial energy 100~GeV traversing a  static medium with temperature 300~MeV for a period of time 4~fm/$c$. The contributions to EEC from medium response and medium-induced radiation are shown in the insets.}
\label{fig2}
\end{figure}

Using the LBT model, we first examine the EEC of a jet produced from a single high-energy  quark interacting with a static medium via elastic and inelastic scatterings. Here, a quark starts with energy $E=100$~GeV and propagates through a medium with temperature $T=300$~MeV for a period of time 4~fm/$c$. The numerical result is shown in Fig.~\ref{fig2}, which compares jet EEC of different flavors and various contributions. In the main figure, a clear flavor hierarchy is observed in both the magnitude and peak position of the EEC due to the quark mass effect: for heavy quarks, there is less medium-induced radiation and the radiation has larger angle with respect to the parent parton. In the insets, we show the flavor hierarchy of the EEC for elastic and inelastic processes separately. Here, the recoil and negative partons produced from elastic scatterings, together with their offsprings, are categorized as ``medium response", while gluons emitted from inelastic scatterings, together with their offsprings, are categorized as ``radiated gluons". In both insets, one also observes smaller EEC for heavy quark jets at small angle due to the mass effect. Interestingly, the EEC for heavy quark jets is close to zero below the dead-cone angle $\theta_0 \sim m_Q/E$ (0.013 for charm and 0.042 for bottom quarks).

{\it \color{blue} Heavy and light flavor jet EEC's in $AA$ collisions --} To study jets in heavy-ion collisions, we sample their production locations using Monte-Carlo Glauber model~\cite{Miller:2007ri}. Each jet parton developed from vacuum shower in Pythia is assumed to stream freely before reaching its formation time (sum of its preceding splitting times)~\cite{Zhang:2022ctd} or the initial time of the QGP (0.6~fm/$c$), whichever is later. Then, it starts interacting with the QGP as simulated by LBT, where Eq.~(\ref{eq:Boltzmann}) is solved in the local rest frame of the medium, with the flow velocity and temperature information provided by a CLVisc hydrodynamics simulation~\cite{Pang:2018zzo,Wu:2021fjf}. The interaction between jet partons and the QGP ceases once the local temperature drops below 165~MeV. In the end, the final-state partons are sorted by minimizing the distance $\Delta R = \sqrt{(\Delta \eta)^2+(\Delta \phi)^2}$ between neighboring partons in the momentum space (where $\eta$ is pseudorapidity and $\phi$ is azimuthal angle), and then connected into strings before feeding back to Pythia for fragmentation to hadrons~\cite{JETSCAPE:2019udz,Zhao:2020wcd,Zhao:2021vmu}. Hadronization of positive and negative partons are conducted separately. Jets in A+A collisions are clustered in the same way as in $p+p$ collisions, except that the Fastjet package has been modified to subtract the momenta of negative particles from those of regular ones~\cite{He:2018xjv}.

\begin{figure}[tbp!]
\includegraphics[width=0.99\linewidth]{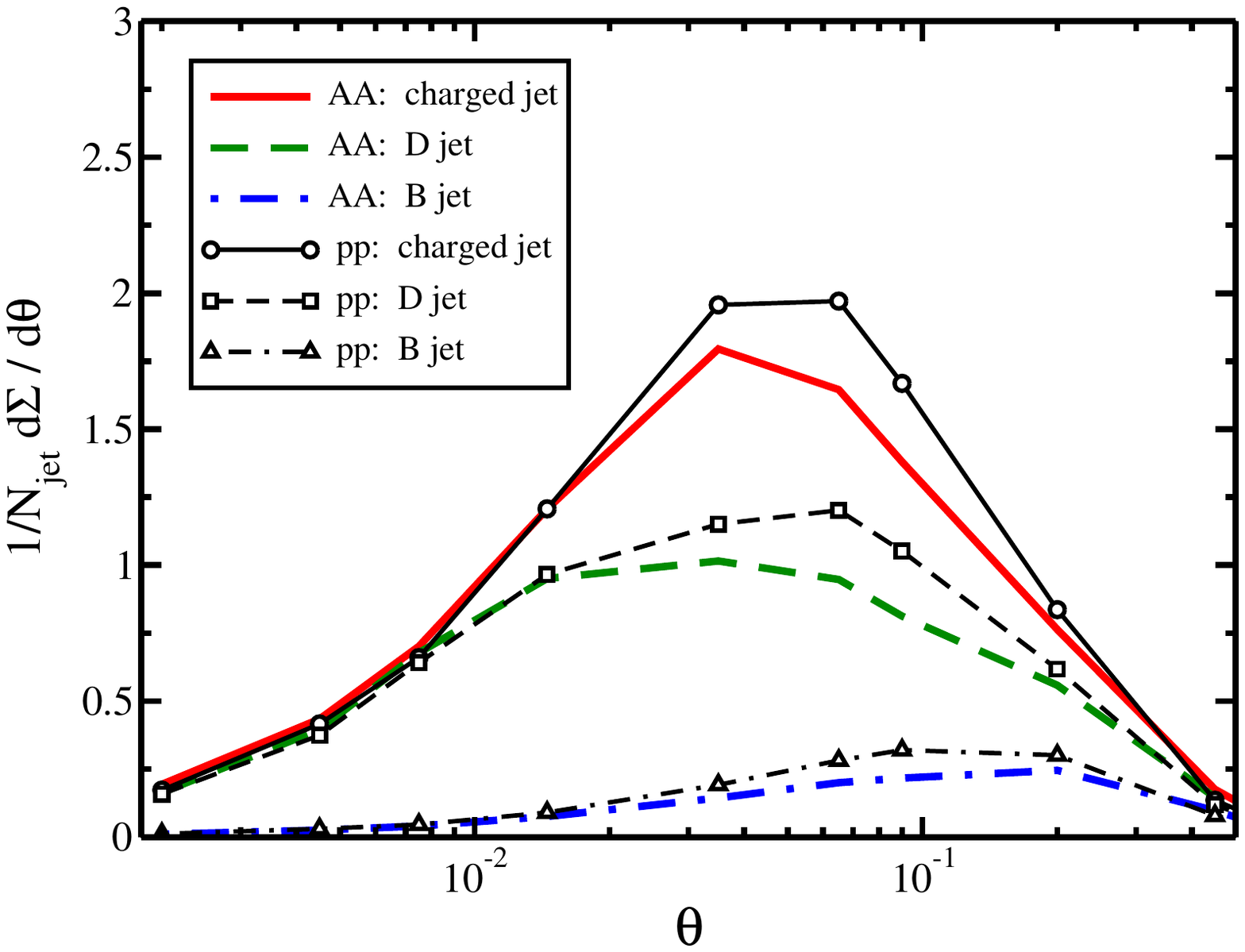}
\includegraphics[width=0.99\linewidth]{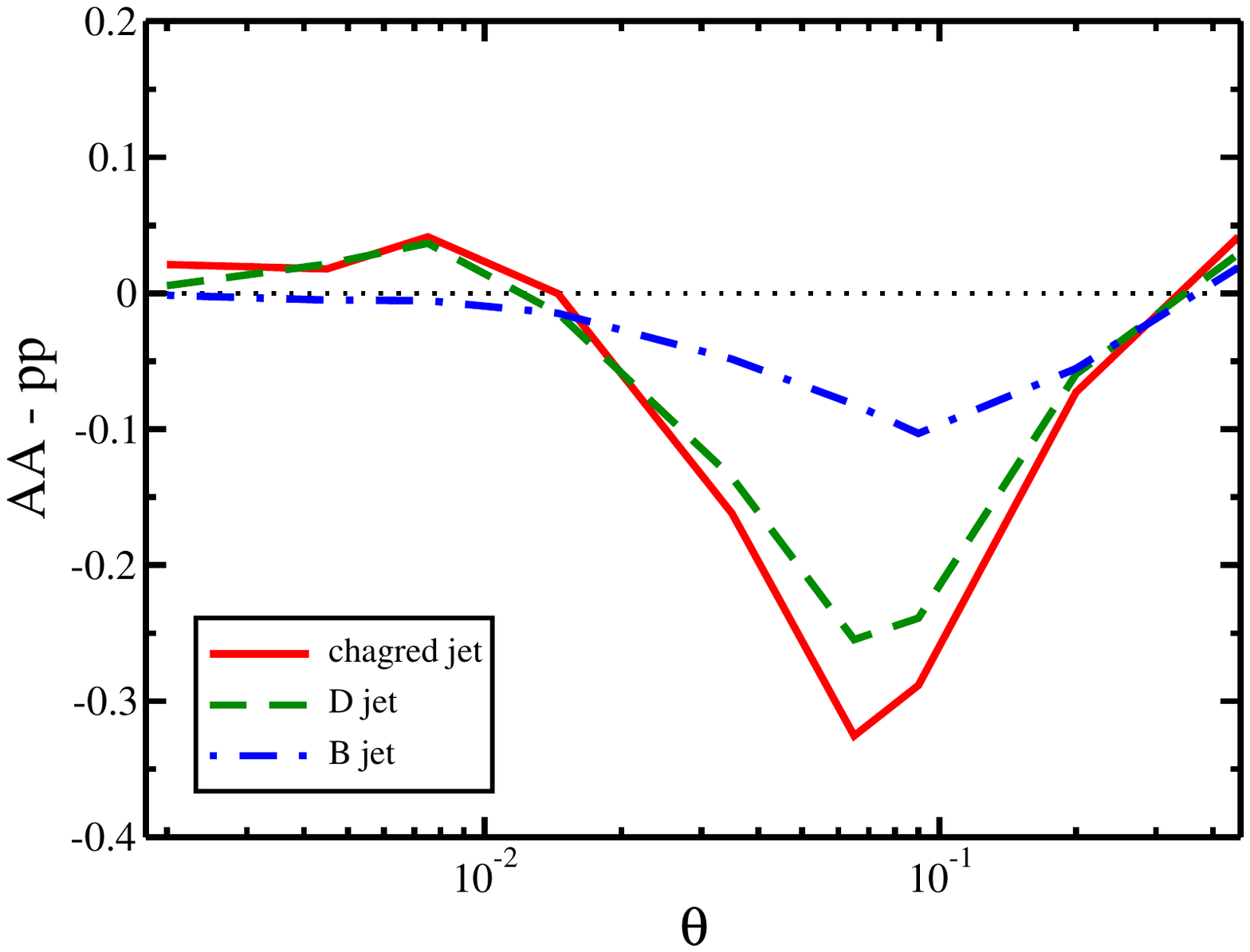}
\caption{(Color online) Upper panel: the EEC of charged hadron jets, $D$-jets, and $B$-jets with $R=0.4$, $p_\mathrm{T}^{\mathrm{jet}} > 40$~GeV and $|\eta^{\mathrm{jet}}| < 0.5$ in $p+p$ and central 0-10\% Pb+Pb collision at $\sqrt{s_\mathrm{NN}}=5.02$~TeV. Lower panel: the difference of the EEC's between Pb+Pb and $p+p$ collisions.}
\label{fig3}
\end{figure}

In Fig.~\ref{fig3}, we show the EEC's for charged hadron jets, $D$-jets, and $B$-jets with $p_\mathrm{T}^{\mathrm{jet}} > 40$~GeV. The upper panel shows the results for $p+p$ and central 0-10\% Pb+Pb collisions at $\sqrt{s_\mathrm{NN}}=5.02$~TeV; the lower panel shows their difference.
One can observe a clear flavor hierarchy for jet EEC's:  for jets tagged by heavier meson, the overall magnitude of EEC is smaller and the angular distribution of EEC also shifts to larger angle.
Comparing $A+A$ to $p+p$ collisions (shown in the lower panel), one can see a strong suppression of EEC at intermediate angle for both heavy and light flavor jets; in the mean time, some mild enhancement at small and large angles is observed.
The suppression at intermediate angles is mainly caused by the energy loss of jet partons inside the QGP, which tends to reduce the hadron production around this typical angle.
Comparing different flavors of jets, we observe that both the suppression (at intermediate angle) and the enhancement (at large and small angles) of EEC become weaker for jets tagged by heavier mesons, indicating that the interaction between heavy quarks and the QGP is weaker due to the mass effect, similar to the result shown Fig.~\ref{fig2} for a simplified environment.
Note no enhancement at small angle for $B$-jet EEC in the current kinematics due to strong mass effect.
The position of the minimum (the largest suppression) in the lower panel also shifts towards larger angles from light flavor jets to $D$-jets and to $B$-jets.

\begin{figure}[tbp!]
\includegraphics[width=0.90\linewidth]{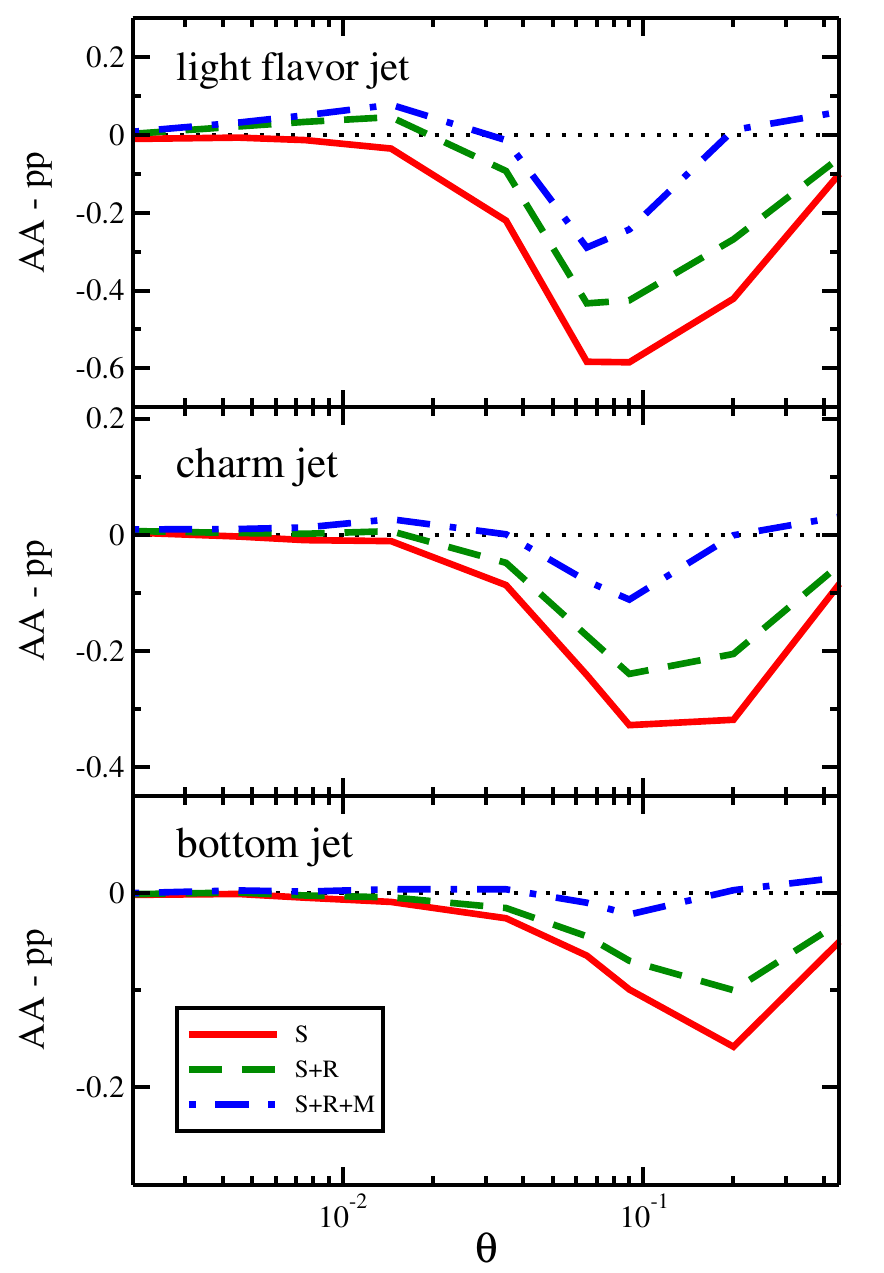}
\caption{(Color online) Contributions from different sources of jet components -- shower partons (S), medium-induced radiations (R), and medium response (M) -- to medium modification of the EEC in central 0-10\% Pb+Pb collisions at $\sqrt{s_\mathrm{NN}}=5.02$~TeV, with the upper panel for light flavor jets, middle for $D$-jets, and lower for $B$-jets. Jets, with $R=0.4$, $p_{\mathrm{T}}^{\mathrm{jet}} > 40$~GeV, and $|\eta^{\mathrm{jet}}| < 0.5$, are constructed using partons with $p_\mathrm{T}>1$~GeV. }
\label{fig4}
\end{figure}

To disentangle the above rich structure of the medium modification of jet EEC, in Fig.~\ref{fig4} we show various contributions -- jet energy loss, medium-induced gluon emission and medium response -- to the EEC for light flavor jets (upper panel), $D$-jets (middle panel), and $B$-jets (lower panel).
Since hadrons are produced from strings that connect partons from all sources, such decomposition can only be done at the partonic level.
Here we construct jets using partons just before hadronization and apply the same kinematics cuts as the hadronic jets.
One first observes that the EEC between shower partons -- partons inherited from Pythia vacuum shower -- is suppressed across the entire range of angles, reflecting parton energy loss inside the QGP.
For jets tagged by heavy quarks, the overall medium modification becomes smaller, and the strongest modification also shifts towards larger angle, both due to the mass effect.
When the contribution from medium-induced radiations is included, the EEC in A+A collisions becomes significantly larger.
In particular, an enhancement of EEC in A+A with respect to $p+p$ collisions is observed at small angles for light flavor jets.
Such enhancement is less pronounced for heavy flavor jets, due to the reduced medium-induced radiation at small angles.
The inclusion of medium response further increases jet EEC, especially at larger angles, reflecting the transport of jet energy from small to large angles via jet-medium interactions.
With all contributions taken into account, a slight enhancement in EEC occurs at large angle for both heavy and light flavor jets.

\begin{figure}[tbp!]
\includegraphics[width=0.95\linewidth]{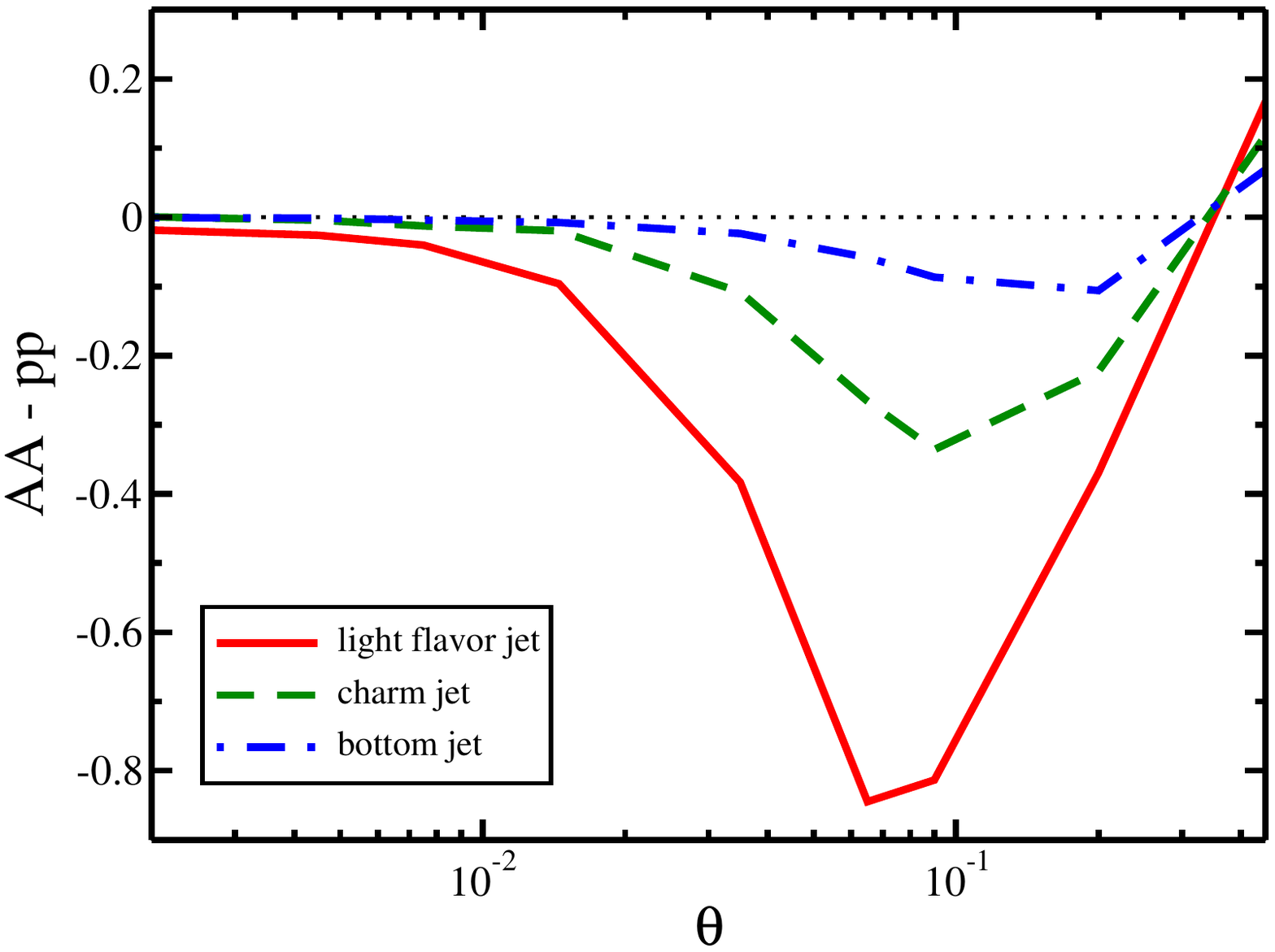}
\caption{(Color online) Nuclear modification of EEC for jets with $R=0.4$, $p_{\mathrm{T}}^{\mathrm{jet}} > 40$~GeV and $|\eta^{\mathrm{jet}}| < 0.5$ in central 0-10\% Pb+Pb collisions at $\sqrt{s_\mathrm{NN}}=5.02$~TeV, with the trigger bias effect excluded.}
\label{fig5}
\end{figure}

To further elucidate the enhancement of jet EEC at small angles, we study the trigger bias effect (due to jet energy loss) on the EEC of inclusive jets.
Similar to Fig.~\ref{fig4}, jets in $p+p$ collisions are constructed using final state partons from Pythia shower.
Partons belonging to vacuum jets with $R=0.4$, $p_{\mathrm{T}}^{\mathrm{jet}} > 40$~GeV, and $|\eta^{\mathrm{jet}}| < 0.5$ are fed to LBT on the jet-by-jet basis.
Without performing jet clustering and selection procedure, the EEC of the final state partons from LBT is analyzed, and the difference from its $p+p$ baseline is presented in Fig.~\ref{fig5}.
Here, the $p_\mathrm{T}$ of vacuum jets is taken as the denominator in Eq.~(\ref{eq:defEEC}) for both vacuum and medium-modified jets.
One observes that previous enhancement of the EEC at small angle in Fig.~\ref{fig4} disappears here in Fig.~\ref{fig5}.
Meanwhile, both the suppression at intermediate angle and the enhancement at large angle become stronger in Fig.~\ref{fig5} compared to Fig.~\ref{fig4}.
These differences result from the trigger bias effect caused by jet energy loss in jet measurement: with the same $p_\mathrm{T}^\mathrm{jet}$ cut, jets experiencing energy loss in A+A collisions have higher initial energy at the production time than those in $p+p$ collisions.
Typically, jets with higher energies tend to radiate more partons with narrower angular distributions, which facilitates the enhancement of the EEC by medium-induced radiation at small angle in A+A collisions, but mitigates the enhancement by medium response at large angle.
In addition, more radiations can offset the reduction of the EEC at intermediate angle by jet energy loss inside the QGP.
Since such trigger bias effect is absent in $\gamma$-triggered jets, we expect that the results shown in Fig.~\ref{fig5} is close to the EEC for $\gamma$-jets.
Qualitative conclusions on the flavor hierarchy of the EEC's at intermediate and large angles should be the same for inclusive jets and $\gamma$-jets.

{\it \color{blue} Summary --} Jet substructure plays essential roles in research fields related to high-energy collider physics, such as search for new physics, precision studies of QCD and probing the dynamics of the QGP.
Heavy quarks provide direct access to probe the mass effect on jet substructure and jet-medium interaction.
The EEC is an extremely interesting observable for studying energy flow within the jets due to its strong sensitivity to jet physics at various scales, either intrinsic or emergent.
In this work, we perform a complete simulation of jet production and medium modification in relativistic heavy-ion collisions, and study the EEC for both heavy and light flavor jets.
A strong flavor hierarchy is observed for the angular distribution of jet EEC: the magnitude is smaller while the typical angle is larger for jets produced from heavy quarks.
The medium modification of inclusive jet EEC in $A+A$ collisions relative to $p+p$ collisions exhibits a very rich structure: suppression at intermediate angle whereas enhancement at large and small angles.
We further disentangle various contributions to jet EEC and find that jet energy loss is responsible for the suppression at intermediate angle, while medium response provide the most significant contribution to the enhancement at large angle.
Another interesting factor is the trigger bias in jet measurement (caused by jet energy loss) which can explain the enhancement of inclusive jet EEC at small angles. Such effect can be verified by comparing the EEC for inclusive jets and $\gamma$-jets; the later does not suffer the trigger bias effect.
Future experiments on jet EEC of different flavors will provide unprecedented opportunities to probe jet physics and the dynamics of QGP at various scales.

{\it \color{blue} Acknowledgments --} We are grateful to Yaxian Mao, Zhong Yang and Wenbin Zhao for helpful discussions. This work is supported in part by the China Postdoctoral Science Foundation under Grant No. 2023M742099 (WJX), and in part by the National Natural Science Foundation of China (NSFC) under Grant Nos.~12175122, 2021-867 (SC, WJX), 12225503 (GYQ) and 11935007 (GYQ, XNW).  Some of the calculations were performed in the Nuclear Science Computing Center at Central China Normal University (NSC$^3$), Wuhan, Hubei, China.

\bibliographystyle{h-physrev5}
\bibliography{WJrefs}

\end{document}